\documentclass[11pt,a4paper]{article}
\usepackage{graphicx}
\usepackage{jcappub}

\title{Impact of the Dark Matter Velocity Distribution on Capture Rates in the Sun}

\author[a]{K.~Choi} 
\author[b]{C.~Rott}
\author[a,c]{Y.~Itow}
\affiliation[a]{Solar-Terrestrial Environment Laboratory, Nagoya University,\\
Furo-cho, Chikusa-ku, Nagoya, 464-8601, Japan}
\affiliation[b]{Department of Physics, Sungkyunkwan University, Suwon 440-746, Korea}
\affiliation[c]{Kobayashi-Maskawa Institute for the Origin of Particles and the Universe, \\
Nagoya University, Furo-cho, Chikusa-ku, Nagoya, 464-8601, Japan}
\emailAdd{koun@stelab.nagoya-u.ac.jp}
\emailAdd{rott@skku.edu}
\emailAdd{itow@stelab.nagoya-u.ac.jp}

\abstract{
Dark matter could be captured in the Sun and self-annihilate, giving rise to an observable neutrino flux. Indirect searches for dark matter looking for this signal with neutrino telescopes have resulted in tight constraints on the interaction cross-section of dark matter with ordinary matter. We investigate how robust limits are against astro-physical uncertainties. We study the effect of the velocity distribution of dark matter in our Galaxy on capture rates in the Sun. We investigate four sources of uncertainties: orbital speed of the Sun, escape velocity of dark matter from the halo, dark matter velocity distribution functions and existence of a dark disc. 
We find that even extreme cases currently discussed do not decrease the sensitivity of indirect detection significantly because the capture is achieved over a broad range of  the velocity distribution by integration over the velocity distribution. The effect of the uncertainty in the high-velocity tail of dark matter halo is very marginal as the capture process is rather inefficient at this region. The difference in capture rate in the Sun for various scenarios is compared to the expected change in event rates for direct detection.
The possibility of co-rotating structure with the Sun can largely boost the signal and hence makes the interpretation of indirect detection conservative compared to direct detection.
}

\notoc


\begin{document}
\maketitle
\flushbottom
\date{\today}

\section{Introduction}

Weakly interacting massive particles~(WIMPs) are prominent candidates to explain the dark matter observed in the Universe. WIMPs (denoted $\chi$) naturally arise in supersymmetry~\cite{Martin:1997ns} and models with large extra dimensions~\cite{Appelquist:2000nn}. They are predicted to have masses ranging from a few GeV to a few TeV~\cite{Bertone:2004pz}. A very promising way to identify the WIMP nature of dark matter is to search for an excess neutrino flux from the Sun generated by self-annihilations of WIMPs~\cite{Press:1985ug,Silk:1985ax,Gould:1987ir,Gould:1991hx} inside the Sun. The high-energy neutrino flux originating from decay of annihilation products as well as a low-energy neutrino signal from hadronic particle showers in the Sun~\cite{Rott:2012qb,Bernal:2012qh} are expected to arrive at neutrino telescopes.
Tight limits on the WIMP-induced neutrino flux from the Sun have been placed with neutrino telescopes such as Super-Kamiokande~\cite{Tanaka:2011uf}, IceCube~\cite{Aartsen:2012kia}, Baksan~\cite{Boliev:2013ai} and ANTARES~\cite{Adrian-Martinez:2013ayv}.
Searches for these neutrinos have to consider various sources of uncertainties on the expected event rates at detectors which can be grouped into three uncorrelated categories:
(1)~uncertainties on annihilation in the Sun,
(2)~uncertainties on neutrino propagation from the Sun to the detector,
(3)~uncertainties related to detector response.
The last two categories are detector- and location-specific and hence need to be studied by the corresponding experiments. The DarkSUSY~\cite{Gondolo:2004sc} / WimpSim~\cite{Blennow:2007tw} packages can, for example, be utilised to obtain precise neutrino flux at the detector.

Uncertainties on the annihilation rate in the Sun are common to searches at all neutrino detectors. For the commonly assumed case of equilibrium between the capture rate and the annihilation rate, the uncertainty on the annihilation is directly given by the uncertainty on the WIMP capture in the Sun. This capture mechanism is theoretically well understood so that one can study the impact of the various effects that contribute to its uncertainty, which is the subject of this work.
Uncertainties on the capture of WIMPs in the Sun originate from a variety of effects. These are related to the WIMP mass, the WIMP-nucleus elastic-scattering cross-section, the local dark matter density, the velocity distribution of dark matter and the composition of the Sun, etc.. Uncertainties coming from the form factor in the scattering cross-section is negligible for spin-dependent (SD) coupling since capture is dominated by hydrogen. For spin-independent (SI) coupling WIMPs, it is predicted to be a few to 20 \% depending on the model used~\cite{Duda:2006uk,Wikstrom:2009kw,Ellis:2009ka}. The uncertainty on the composition of the Sun is predicted to be negligible for SD coupling~\cite{Wikstrom:2009kw} and at the 20\% level for SI coupling WIMPs~\cite{Ellis:2009ka}.

Larger uncertainty comes from the understanding of the local dark matter phase space structure. The uncertainty in local dark matter density is expected to be a factor of two~\cite{Catena:2009mf,Weber:2009pt,Salucci:2010qr,Pato:2010yq,Garbari:2011dh}, as are the uncertainties in capture rate and signal rate of direct searches, as scattering rate simply increases with the local density. However, the local velocity distribution of dark matter impacts direct and indirect searches very differently. As opposed to direct detection looking for energetic scattering of WIMPs, less energetic WIMPs are more easily captured by the Sun. 
Therefore direct detection is sensitive to WIMPs with high-velocity, while indirect detection is sensitive to the low part of the velocity distribution. In the literature there have been extensive discussions about the impact of uncertainties related to WIMP phase space on direct detection~\cite{Green:2002ht,Fairbairn:2008gz,MarchRussell:2008dy,Strigari:2009zb,
Kuhlen:2009vh,Cerdeno:2010jj,McCabe:2010zh,Green:2010gw,Arina:2011si,Green:2011bv,
Fairbairn:2012zs,Strigari:2012gn,Bozorgnia:2013pua,Peter:2013aha,Fornasa:2013iaa} and indirect detection for more specific scenarios~\cite{bruch:2009rp,Peter:2009qj,Ling:2009cn,Serpico:2010ae,Kundu:2011ek,Rott:2011fh,Sivertsson:2012qj,Arina:2013jya}.

In this work, we will give a comprehensive and general treatment of uncertainties from the dark matter velocity distribution on indirect solar WIMP searches. Previous works have only discussed specific effects, while this work quantifies effects as function of the WIMP mass, making it easy to incorporate in future experimental and theoretical works.
This paper is structured as follows: 
In section~\ref{sec:Capture}, we first review the WIMP capture process in the Sun to manifest how the dark matter velocity distribution factors into it. In section~\ref{sec:vdf}, we introduce a variety of velocity distributions, for which we then discuss the impact on the capture rate in section~\ref{sec:uncertainty}. In section~\ref{sec:discussion}, we relate our finding to velocity distribution dependencies for direct detection experiments. We conclude and summarise our results in the final section~\ref{sec:Conclusions}.

\section{WIMP capture and annihilation in the Sun}
\label{sec:Capture}

In this section we review the WIMP capture mechanism that leads to the accumulation of WIMPs in the Sun. The capture process has been extensively discussed in the literature~\cite{Gould:1987ir,Gould:1991hx}, which will be reviewed here to see in particular how the dark matter velocity distribution is involved. 

WIMPs which are abundant in the dark matter halo of the Milky Way can scatter off a nucleus in the Sun as it revolves around the Galactic centre. The WIMP may lose enough energy by elastically scattering off a nucleus to fall below the escape velocity of the Sun at the point of scatter to be gravitationally captured. The differential WIMP capture rate on nucleus $i$ per unit shell volume $dV$ at distance $r$ from the centre takes the following form :
\begin{equation}\label{eq:capture}
\frac{dC_{i}}{dV}= \frac{\rho_{H}}{M_{\chi}}\int_{0}^{\infty}du
\frac{f_{\eta}(u)}{u}\Omega_i(Q),
\end{equation}
where $\rho_{H}$ is the local dark matter halo density set to be 0.3~{\rm GeV/cm$^3$}~\cite{Kamionkowski:1997xg,Yao:2006px} and $M_{\chi}$ is the mass of the WIMP.
$f_{\eta}(u)$ is the velocity distribution function~(VDF) seen in the reference frame of the Sun. For isotropic VDF, $f_{\eta}(u)$ can be simply calculated from the original VDF in the Galactic frame, $f_o(v)$, as
\begin{equation}
\label{eq:convert}
f_{\eta}(u) = \int^{1}_{-1}f_o(\sqrt{v^2+v_{\odot}^2+2 v v_{\odot}cos\theta})~dcos\theta,
\end{equation}
where $\vec{u}=\vec{v}+\vec{v_{\odot}}$ and $\theta$ is the angle between $\vec{v}$ and $\vec{v}_{\odot}$.
The bound velocity of a WIMP in the gravitational field of the Sun, $w$, and the escape velocity at scattered position, $u_{esc}(r)$ are related via $w^2 = u^2 + u_{esc}(r)^2$. A WIMP can be captured when $w$ drops below $u_{esc}(r)$ after losing kinetic energy through scattering off a nuclei. The capture probability $\Omega_i(Q)$ can be written as~\cite{Gould:1987ir}
\begin{equation}
\Omega_i(Q)=\sigma_{i} n_i\frac{M_i}{2\mu_i^2} \int^{Q_{max}}_{Q_{min}} F^2(Q) dQ,
\end{equation}
where $\sigma_{i}$ is the WIMP-nucleus $i$ elastic-scattering cross-section at zero-momentum transfer, $n_{i}$ is the number density of nucleus $i$ in the Sun, $M_i$ is the mass of nucleus $i$, $\mu_i$ is the reduced mass of nucleus $i$ and the WIMP, $Q$ is the recoil energy where
\begin{equation}\label{eq:qmin}
Q_{\rm min}=\frac{1}{2}M_{\chi} u^2
\end{equation}
is the minimum recoil energy required to be captured,
\begin{equation}\label{eq:qmax}
Q_{\rm max}=\frac{1}{2}\beta_{+}M_{\chi}w^{2},
\end{equation}
is the kinematically determined maximal recoil energy with 
\begin{equation}
\beta_{\pm} = \frac{4M_i M_{\chi}}{(M_i \pm M_{\chi})^2}
\end{equation}
and $F^2(Q)$ is a nuclear form factor taken to have exponential form from~\cite{Gould:1987ir}.
For SD WIMP capture we neglect the contributions from the heavier elements than hydrogen. For SI coupling case, we assume isospin conserving scattering and write the SI zero-momentum transfer cross-section to a nucleus $\sigma^{SI}_{i}$ with the mass number $I$ as
\begin{equation}\label{sigma}
\sigma^{SI}_{i} = \sigma^{SI}_p I^2 \frac{\mu_i^2}{\mu_p^2}
\end{equation}
where $\sigma^{SI}_p$ is the SI WIMP-nucleon zero-momentum transfer scattering cross-section and $\mu_{p}$ is the WIMP-nucleon reduced mass.

To calculate the total capture rate, the above expression has to be summed over all relevant nuclear species inside the Sun and integrated over the distance $r$ from the centre as
\begin{equation}
 C_C = \int_{0}^{R_{\odot}} 4 \pi r^2 dr \sum_i \frac{dC_i}{dV}, 
\label{eq:caps8}
\end{equation}
where $R_{\odot}$ is the radius of the Sun.

The maximum velocity of a WIMP to be captured by nucleus $i$ at distance $r$ from the centre of the Sun is given where $Q_{min}$~(eq.~\ref{eq:qmin}) equals to $Q_{max}$~(eq.~\ref{eq:qmax}), where
\begin{equation}u_{max}(i)=\sqrt{\beta_{-}}u_{esc}(r).
\label{eq:umax}
\end{equation}

In the case of $\sqrt{\beta_{-}}\sim1$, the escape velocity at the surface of the Sun ($\sim618~{\rm km/s}$), being larger than the mean velocity of WIMPs in the Galactic halo ($\sim300~{\rm km/s}$), allows efficient capture process. However, when mass-matching between the WIMP mass and the mass of the nucleus fails, $\beta_-$ will get smaller than one, resulting in suppressed capture of high-velocity WIMPs.

\begin{figure*}[t]
  \centering
\includegraphics[width=.9\textwidth]{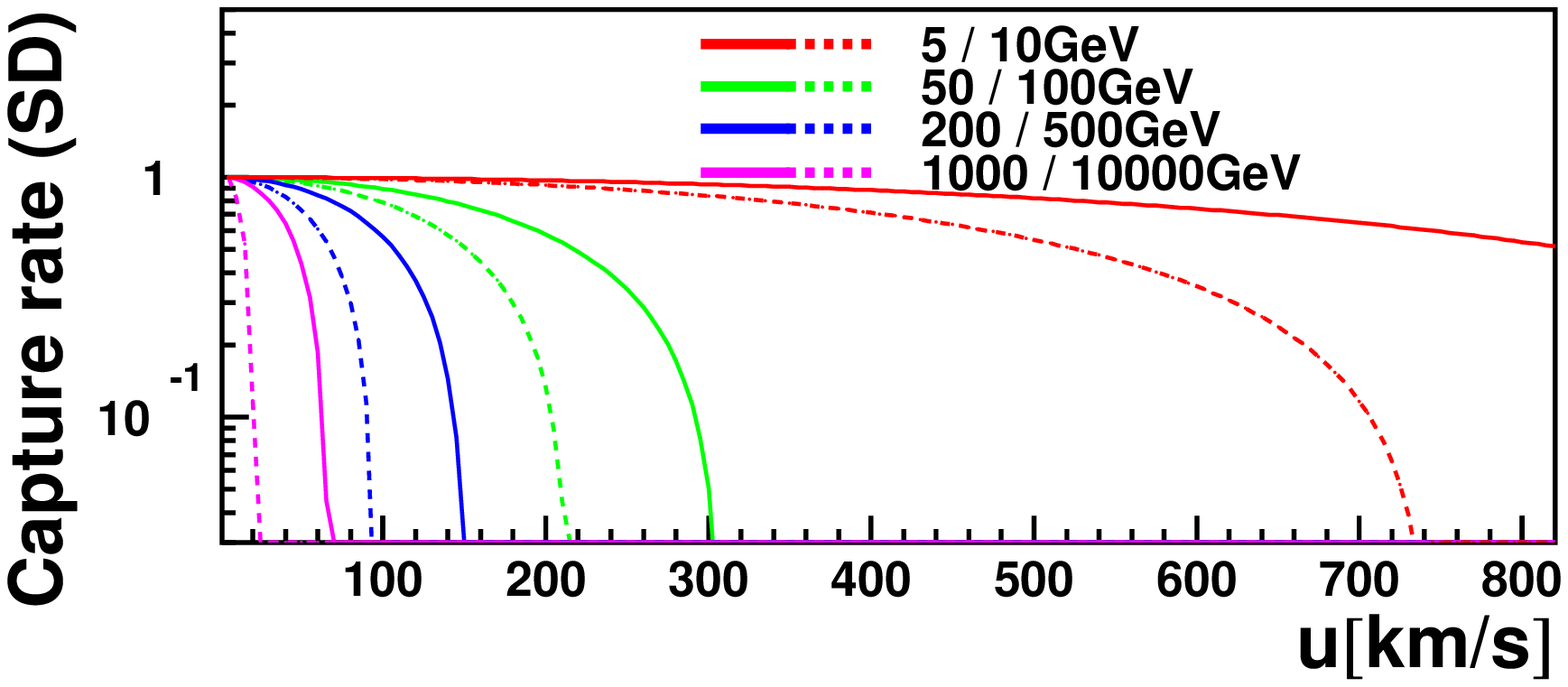}
\includegraphics[width=.9\textwidth]{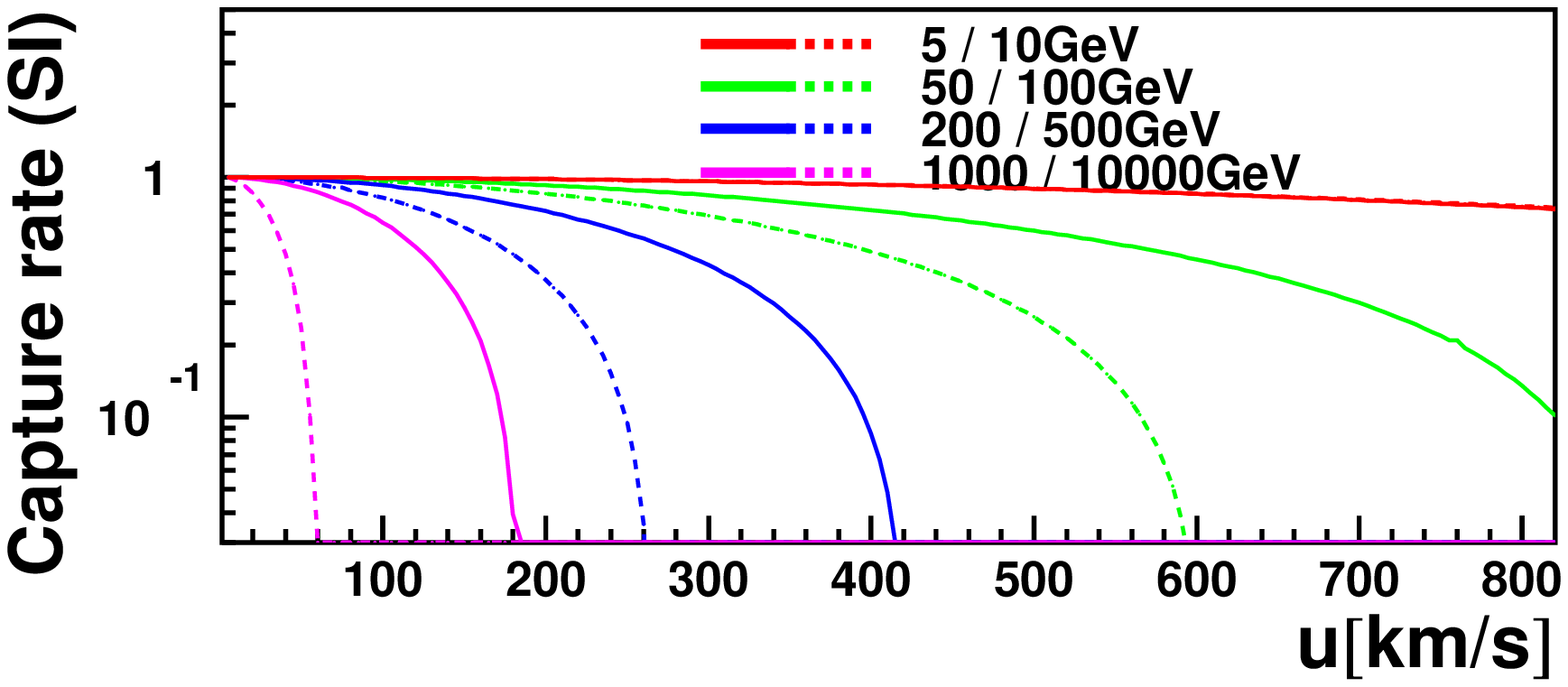}
  \caption{The capture rate for a sample of representative WIMP masses as function of the relative velocity to the Sun is shown. The rates are normalised to the rate at $u=0$. We show the SD case (top) compared to the SI case (bottom), for eight selected masses of 5 GeV (red solid)/10 GeV (red dashed), 50 GeV (green solid)/100 GeV (green dashed), 200 GeV (blue solid)/500 GeV (blue dashed), 1000 GeV (magenta solid)/10000 GeV (magenta dashed).}
\label{fig:fig1}
\end{figure*}

\begin{figure*}[t]
  \centering
\includegraphics[width=.8\textwidth]{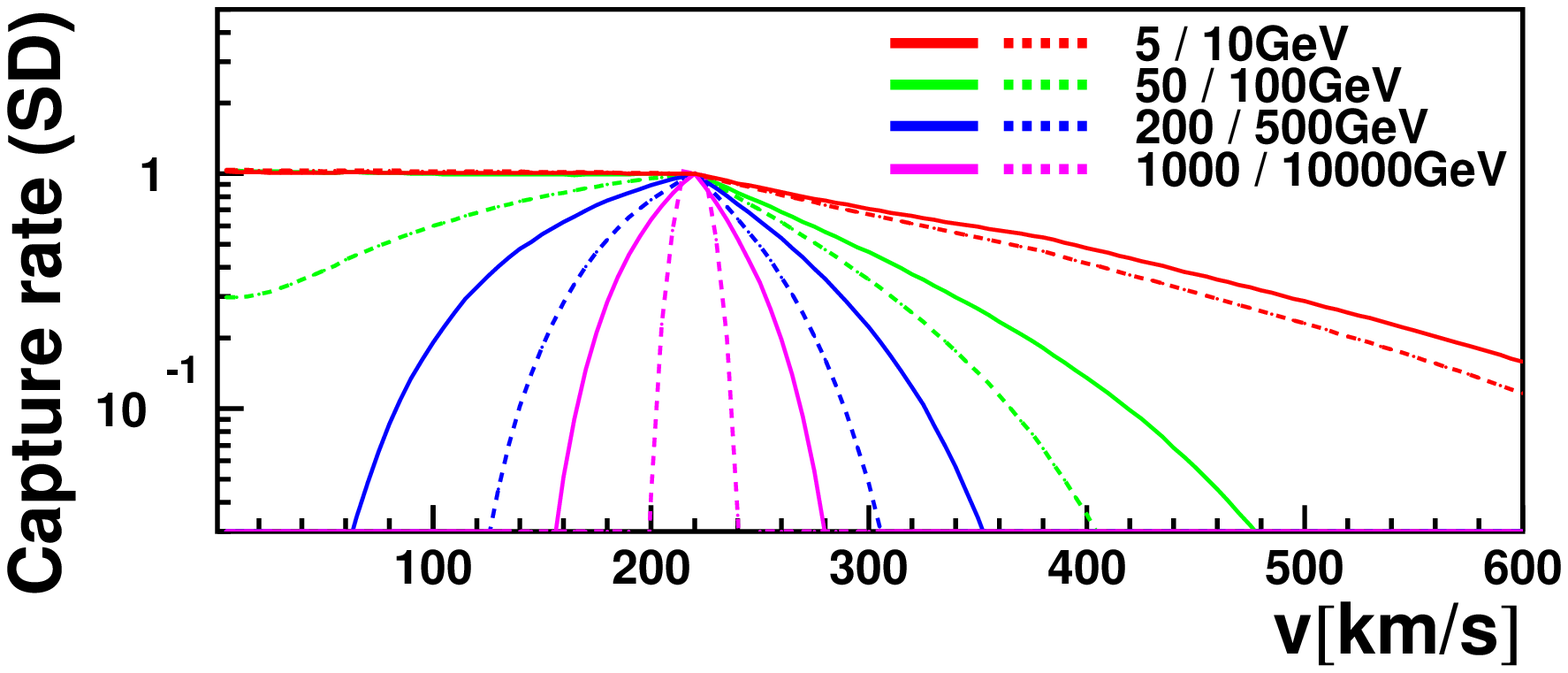}
\includegraphics[width=.8\textwidth]{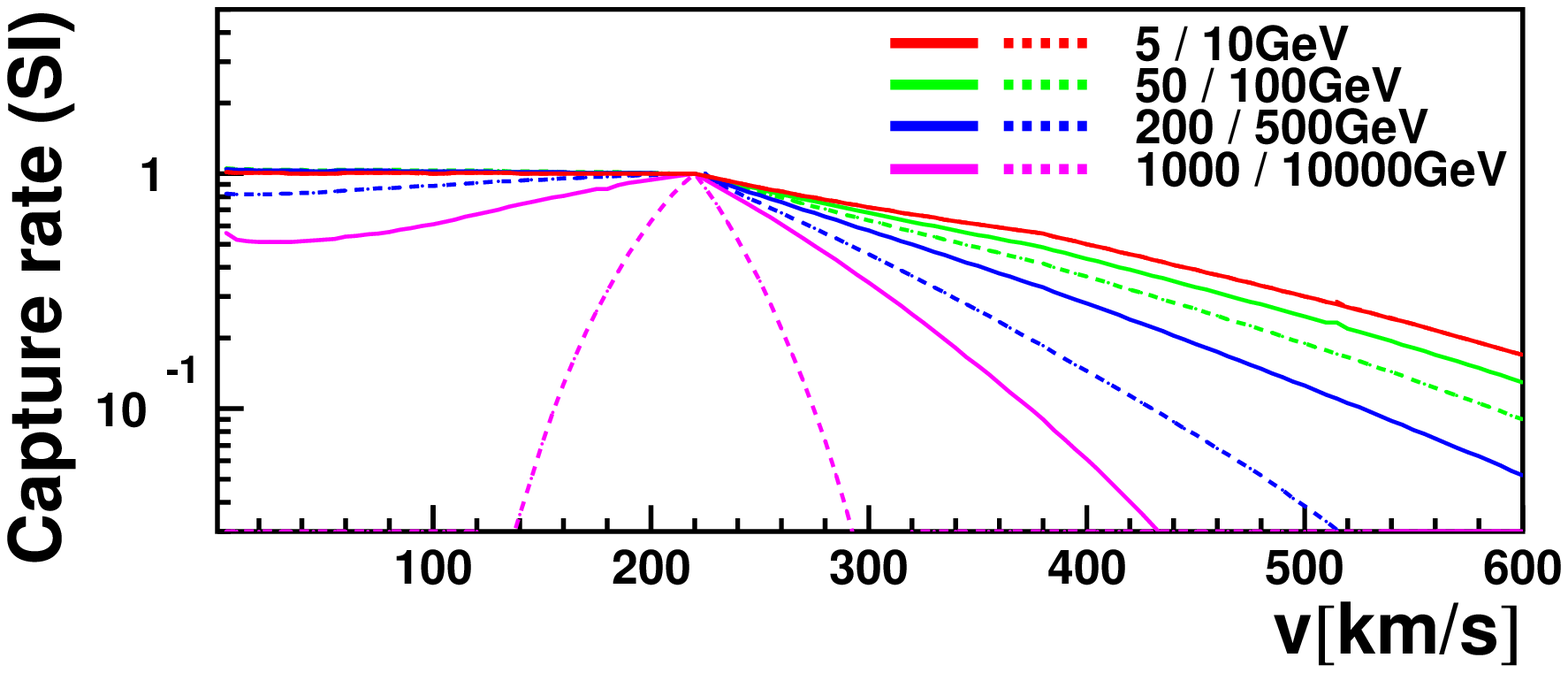}
  \caption{The capture rate for sample WIMP masses as function of the velocity in the Galactic frame is given relative to the capture rate for $v=220~{\rm km/s}$. We show the SD case (top) compared to the SI case (bottom), for eight sample masses of 5 GeV (red solid)/10 GeV (red dashed), 50 GeV (green solid)/100 GeV (green dashed), 200 GeV (blue solid)/500 GeV (blue dashed), 1000 GeV (magenta solid)/10000 GeV (magenta dashed).}
\label{fig:fig2}
\end{figure*}

To estimate $u_{max}$ for the total capture rate, we have to consider contributions from multiple elements with their radial distribution inside the Sun.
Figure~\ref{fig:fig1} and Fig.~\ref{fig:fig2} show the relative capture rate for several WIMP masses and separately for SD and SI capture as function of the dark matter velocity in the local moving frame of the Sun and in the Galactic frame, respectively. Capture rates are calculated using a delta function for the VDF and are normalised to the highest capture rate for each WIMP mass.
In Fig.~\ref{fig:fig1}, it can be seen that the capture efficiency decreases as the velocity increases, which is simply related to the capture kinematics. 
The distribution of the energy loss is uniform over the interval $0 < Q < Q_{max}$ and if the WIMP fails to transfer a recoil energy $Q$ of at least $Q_{min}$ to the nucleus, it will not be captured.
As $u$ increases towards $u_{max}$, $Q_{min}$ increases towards $Q_{max}$, reducing the range of recoil energies which would result in capture and therefore reducing the capture rate. 
Hence, capture will be more efficient for WIMPs with lower velocities.
As the capture in the Sun is dominated by the most abundant hydrogen and other light elements, for light WIMPs good mass-matching will allow the capture to be efficient until the high-velocity tail, whereas for heavy WIMPs it starts losing efficiency earlier. SI capture in comparison to SD capture is efficient for in the entire velocity region up to WIMP masses of about 50~GeV, which is the `resonance range'~\cite{Gould:1987ir,Jungman:1995df}, helped by mass-matching scatterings with heavier elements such as He, O, Si and Fe.

Once captured, WIMPs lose energy in subsequent scatters and settle at the centre of the gravitational well and thermalise. WIMPs can be eventually accumulated in the centre of the Sun in this fashion. 
The trapped WIMPs can then self-annihilate and create standard model particles from which only neutrinos could escape and be detected.
WIMPs could also become unbound from the Sun in scatters, a process known as evaporation. For our study we constrain ourselves to WIMP masses above 4~GeV, a region where evaporation is negligible~\cite{Griest:1986yu,Gould:1987ju,Hooper:2008cf,busoni:2013kaa}.
For a sufficiently large scattering cross-section for the WIMP mass~\cite{Peter:2009mk}, the evolution of the total number of WIMPs in the Sun can be described by the following differential equation
\begin{equation}\label{eq:diff}
\frac{dN}{dt}=C_{C}-C_{A}N^{2}
\end{equation}
where N is the total number of WIMPs in the Sun, $C_{C}$ is the capture rate and $C_{A}$ is linked to the annihilation rate, $\Gamma_{A}=C_{A}N^{2}/2$.
The current annihilation rate is given by
\begin{equation}
\label{gamsun}
\Gamma_{A}=\frac{C_C}{2}\tanh^{2}(t_\odot/\tau).
\end{equation}
Here $t_{\odot}$ is the age of the Sun, $\tau = (C_{C} C_{A})^{-1/2}$ is the time scale of WIMP capture to be in equilibrium with annihilation, also known as the equilibration time.
Given the age of the Sun it is frequently assumed that equilibrium between capture and annihilation has been achieved. 
For this case :
\begin{equation}
\ \Gamma_{\rm A}^{\rm equi} = \frac{1}{2} C_{C}.
\label{eq:E}
\end{equation}
While differential equation~\ref{eq:diff} only holds for the instantaneous thermalization model~\cite{Peter:2009mk}, the annihilation rate at equilibrium is still unchanged as long as the thermalization time is short compared to the age of the Sun.  So, when the capture rate, $C_{\rm C}$, regulates the annihilation rate, $\Gamma_{\rm A}$, as expected, a limit on the neutrino flux sets a limit on the WIMP-nucleon scattering cross-section.

\section{Dark matter velocity distributions}
\label{sec:vdf}
In this section we survey dark matter velocity distributions. We introduce the standard halo model, then VDFs obtained from simulations. We also introduce scenarios with a co-rotation structure (dark disc). These are later used as benchmark scenarios for which we compute capture rates of WIMPs in the Sun.

For a simple isotropic sphere with density profile $\rho(r) \propto r^{-2}$ of collisionless particles, the velocity distribution leads to a so-called Maxwellian:
\begin{equation}
\label{eq:maxv}
f(v)dv= \frac{4}{\sqrt{\pi}}\Bigl(\frac{3}{2}\Bigr)^{\frac{3}{2}} \frac{v^2}{{v_{rms}}^3} \exp\Bigl({-\frac{3}{2}\frac{v^2}{{v_{rms}}^2}}\Bigr)\Theta(v-v_{esc}) dv.
\end{equation}
The high-velocity tail is truncated by the Galactic escape speed $v_{esc}$, for which we use $544~{\rm km/s}$ as a default value and normalise the distribution to unity after the cut. In the reference frame of the Sun it can be written as:
\begin{equation}
\label{eq:maxv}
 f(u) = \sqrt{\frac{3}{2\pi}} \frac{u}{v_{\odot}v_{rms}} 
\Bigl( \exp \Bigl(- \frac{3(u - v_{\odot})^2}{2 v_{rms}^2} \Bigr) - 
\exp \Bigl( - \frac{3(u + v_{\odot})^2}{2 v_{rms}^2} \Bigr) \Bigr).
\end{equation}
The observed value of the orbital speed in the solar position $v_\odot$ is about 220~{\rm km/s}, which leads the local velocity dispersion of the dark matter halo $v_{rms}$ to be $\simeq \sqrt{3/2}v_{\odot}\simeq270~{\rm km/s}$. 

\begin{figure*}
\begin{minipage}[t]{.5\textwidth}
\includegraphics[width=1.\textwidth]{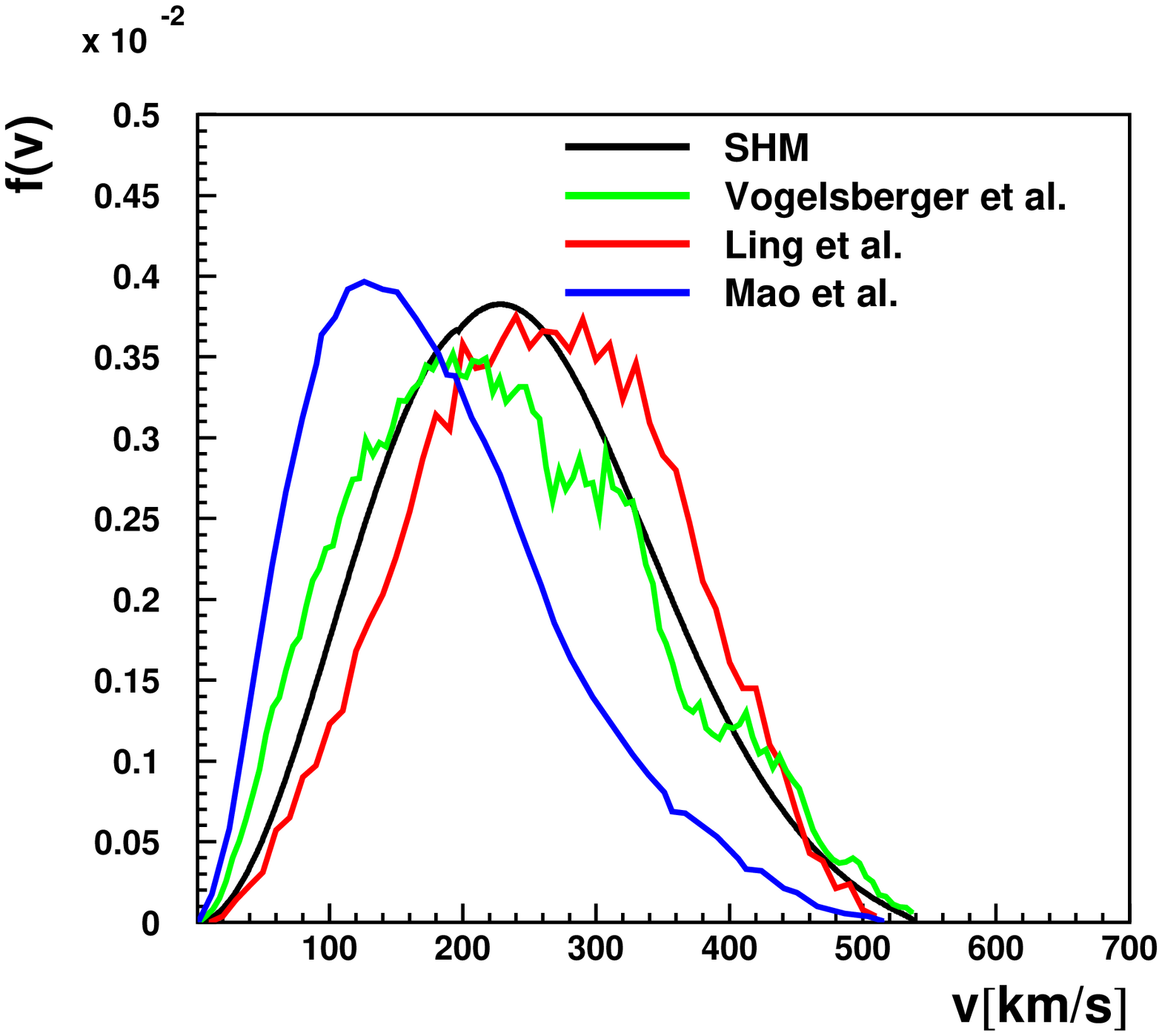}
\end{minipage}
\hfill
\begin{minipage}[t]{.5\textwidth}
\includegraphics[width=1.\textwidth]{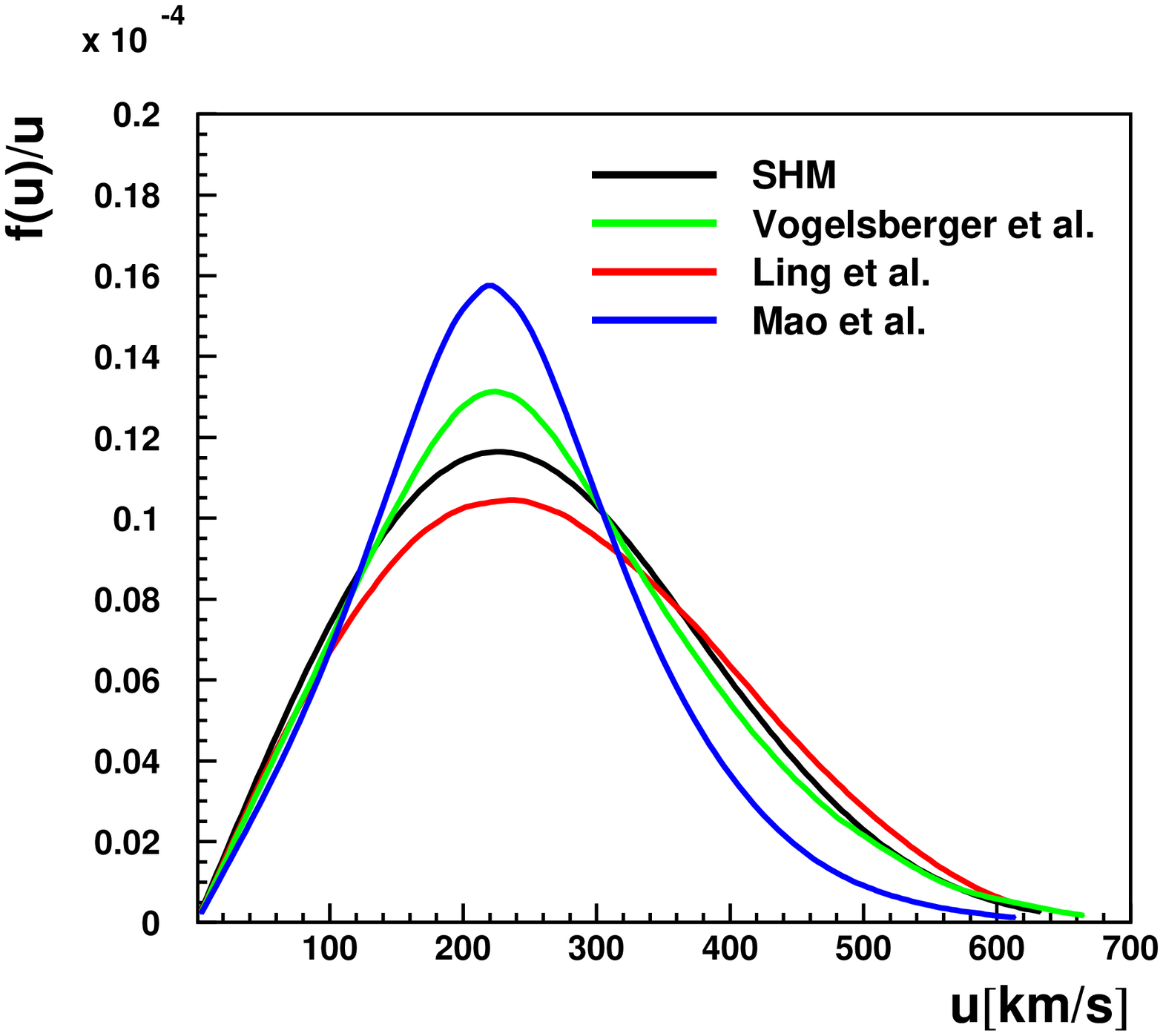}
\end{minipage}
\hfill
\caption{The VDFs of dark matter halo from simulations are shown together with the standard halo model~(SHM). Left panel shows the normalised VDF $f(v)$s in the Galactic frame and right panel shows the VDFs in the local moving frame of the Sun in the form of  $f(u)/u$ in eq.~\ref{eq:capture}. Detailed descriptions for Vogelsberger et al.~\cite{Vogelsberger:2008qb} (green), Ling et al.~\cite{Ling:2009eh} (red), Mao et al.~\cite{Mao:2012hf} (blue) simulated halos are given in the text.}
\label{fig:fig3}
\end{figure*}

We will call the Maxwellian velocity distribution with introduced parameters above `standard halo model~(SHM)' throughout this work. As this model is commonly used, it serves as a benchmark model in our study, however it is know to have theoretical inconsistencies and discrepancies are expected due to Galactic dynamics~\cite{Kundu:2011ek}. 

Recent cold dark matter N-body simulations confirm a significant deviation of the VDF of dark matter halos from the SHM~\cite{Vogelsberger:2008qb,Ling:2009eh,Kuhlen:2009vh,Mao:2012hf}. We choose three benchmark VDFs from recent works (see Fig.~\ref{fig:fig3} left) and convert them using equation \ref{eq:convert} to the local moving frame of the Sun (Fig.~\ref{fig:fig3} right). The small structures seen in the Galactic frame are washed out in the local moving frame of the Sun. VDFs in the local moving frame of the Sun are shown in the form of $f(u)/u$ as this term is the relevant one for the capture rate (see eq.~\ref{eq:capture}). Our benchmark distributions are taken from three recent N-body hydrodynamical simulations: the Aquarius~\cite{Springel:2008cc} project which resolved a Milky Way-sized Galactic halo with more than a billion particles; an N-body simulation with Baryons~\cite{Ling:2009eh} carried with the cosmological Adaptive Mesh Refinement (AMR) code RAMSES~\cite{Teyssier:2001cp}; the Rhapsody cluster re-simulation project~\cite{Wu:2012wu}.

In Fig.~\ref{fig:fig3} the green line shows a VDF obtained with the Aquarius project, taken from Fig.2 in Vogelsberger et al.~\cite{Vogelsberger:2008qb}, which we adopt as one of our benchmark VDFs. It is the median of the velocity modulus distributions for 2~kpc boxes centred between 7 and 9~kpc from the centre of simulated halo, Aq-A-1~\cite{Springel:2008cc}.
The broad structure seen above 250~{\rm km/s} originates from similar features seen in each of the single boxes at approximately the same velocity with similar amplitude. Authors found that these features appear in all of the six simulated halos, which suggests that they do not reflect local structures but rather a global property of the inner halo, possibly a consequence of real dynamic process. We take this VDF as an example for a VDF with a more complex structure.

The velocity modulus in a spherical shell $7 < R < 9$~kpc around the Galactic centre, shown in Fig.4(d) of Ling et al.~\cite{Ling:2009eh} is adopted as our second benchmark VDF and shown as red line in Fig.~\ref{fig:fig3}. The VDF is platykurtic ($K <$ 3), i.e. broader than a Gaussian distribution with the same standard deviation. 
Authors noted that equilibrated self-gravitating collisionless structures can exhibit a Tsallis distribution, which describes non-extensive systems. For the particles in this shell the VDF is indeed well fit by a Tsallis distribution and the best fit kurtosis parameter $K$ = 2.44 agrees with the observation ($K$ = 2.39) better than a best-fit generalised Maxwellian ($K$ = 2.71).
The best-fit Tsallis distribution velocity dispersion parameter $v_0$ is 267.2 km/s. This VDF will serve as a representative of a realistic description of a galaxy considering interactions between the baryonic and the dark components.

Our third benchmark velocity distribution is taken from Fig.1 of Mao et al.~\cite{Mao:2012hf}, which is the stacked VDF for 96 halos in Rhapsody simulations~\cite{Wu:2012wu}. Authors argue that the largest current theoretical uncertainty on the VDF arises from the unknown radial position of the solar system relative to the dark matter halo scale radius, $r/r_s$. In the range of current observation (\cite{Mao:2012hf} and its refs) $r/r_s$ is suggested to be within 0.15 $\sim$ 6. We picked the marginal value of $r/r_s=0.15$ to represent a halo model peaking at low-velocity.

Descriptions of our benchmark VDFs are summarised in table~\ref{Tab:1}.
Note that for simplicity we regard VDFs to be isotropic throughout our discussion.

\begin{table}[htb]
\begin{center}
\begin{tabular}{|| p{3cm} | p{4cm} | p{5cm} | p{1.4cm}||}
\hline
\hline
 Name  & Description & Description  & Source \\
\& Reference  & of simulation & of VDF  &  \\
\hline
Vogelsberger et al.~\cite{Vogelsberger:2008qb} & largest dark matter only simulation of a Milky Way-sized dark matter halo & median of the velocity modulus distributions for 2 kpc boxes centred between 7 and 9 kpc from the Galactic centre, with a broad bump at $v\simeq250~{\rm km/s}$ & Fig.2\\
\hline
Ling et al.~\cite{Ling:2009eh}  & a Milky Way-sized galaxy from high-resolution N-body simulation with baryons & velocity modulus in a spherical shell $7 < R < 9$~kpc around the Galactic centre, with platykurtic shape ($K$ = 2.39) and high velocity dispersion ($v_0$ = 267.2 km/s) & Fig.4(d)\\
\hline
Mao et al.~\cite{Mao:2012hf} & cluster re-simulation project & stacked velocity distribution for 96 halos at $r/r_s=0.15$, peaks at low-velocity & Fig.1\\
\hline
\hline
\end{tabular}
\caption{The sources and descriptions of the benchmark VDFs used in this work are summarised.}\label{Tab:1}
\end{center}
\end{table}

The merger history of the Milky Way favours the existence of a co-rotating structure made from materials accreted from merged satellites, known as dark disc~\cite{Ling:2009eh,Read:2008fh,Read:2009iv,Purcell:2009yp,Kuhlen:2013tra}.
Simulations show that the local density of the dark disc $\rho_{dd}$ could range from a few percent~\cite{Kuhlen:2013tra} up to $\sim$ 1.5~times~\cite{Read:2009iv} of the density of the local dark matter halo, $\rho_{H}$. We consider scenarios without a dark disc and the cases where the local dark disc density is a quarter ($\rho_{dd}/\rho_{H}$=0.25) and for an extreme case equal ($\rho_{dd}/\rho_{H}=1.0$) to the local dark matter halo density. For the dark disc we assume as default a Maxwellian VDF with velocity dispersion $\sigma=50~{\rm km/s}$ and the relative circular speed to the Sun, $v_{lag}=50~{\rm km/s}$ following~\cite{Bruch:2008rx} and~\cite{Ling:2009cn}.

\begin{figure*}
\begin{minipage}[t]{.5\textwidth}
\includegraphics[width=1.\textwidth]{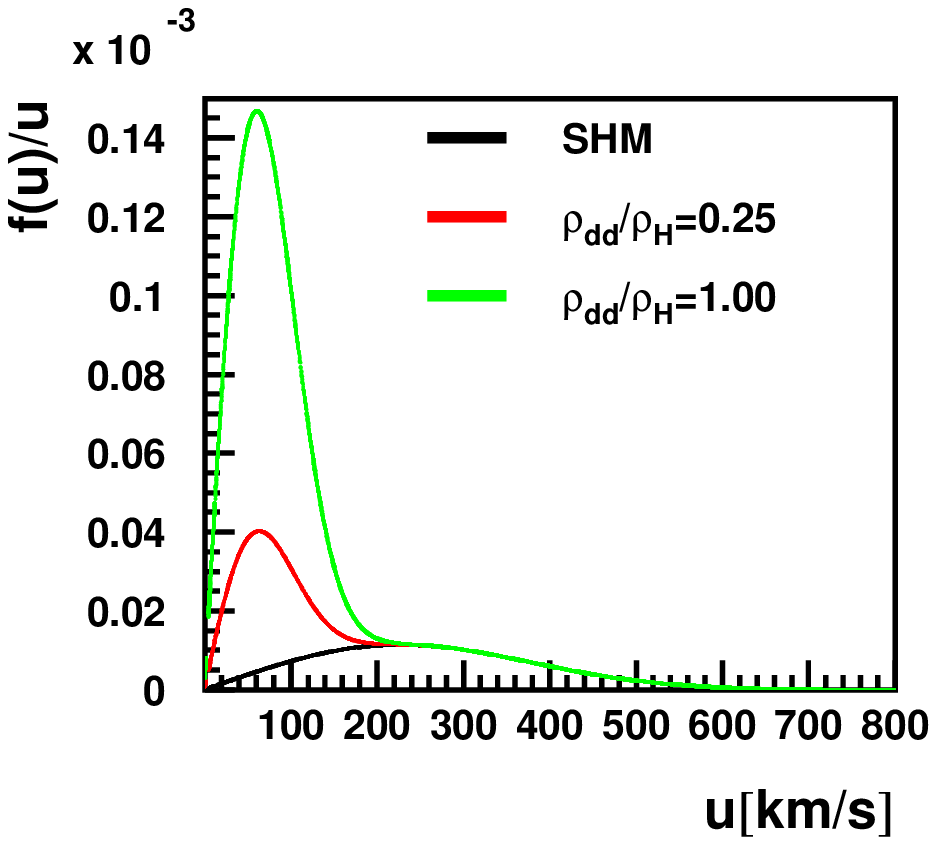}
\end{minipage}
\hfill
\begin{minipage}[t]{.5\textwidth}
\includegraphics[width=1.\textwidth]{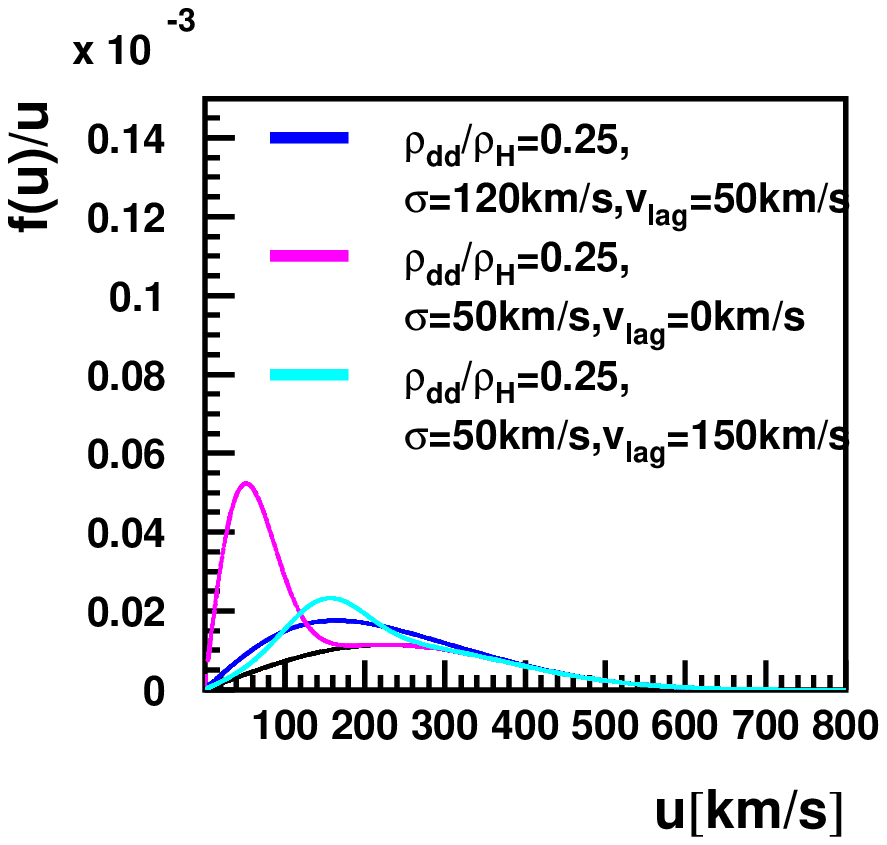}
\end{minipage}
\hfill
\caption{The dark disc velocity distributions added to dark matter halo in the local moving frame of the Sun are shown in the form of $f(u)/u$.
red: with $\rho_{dd}/\rho_{H}=0.25$, $\sigma=50~{\rm km/s}$, $v_{lag}=50~{\rm km/s}$,
green: $\rho_{dd}/\rho_{H}=1$,
blue: $\sigma=120~{\rm km/s}$,
magenta: $v_{lag}=0~{\rm km/s}$,
cyan: $v_{lag}=150~{\rm km/s}$.
Unless otherwise specified in the legend, parameters are fixed to $\rho_{dd}/\rho_{H}=0.25$, $\sigma=50~{\rm km/s}$, $v_{lag}=50~{\rm km/s}$ and $\rho_{H}$ = $0.3~{\rm GeV/cm}^3$ (Note that the total dark matter density is increased to $\rho_{dd} + \rho_{H}$). The SHM with no dark disc is shown for comparison~(black).}
\label{fig:fig4}
\end{figure*}

The dark disc velocity distribution varies for different scenarios, we therefore select a few representative descriptions. Further to cover uncertainties of $v_{lag}$ and the velocity dispersion we sample various cases. Uncertainties on $v_{lag}$ are discussed elsewhere~ \cite{Read:2009iv} and for a discussion about the uncertainty in the velocity dispersion which could potentially be large we refer the reader to~\cite{Purcell:2009yp}. In future more realistic full cosmological hydrodynamics simulations~\cite{Ling:2009cn} might provide more clarity. We here test a variety of dark disc implementations, we varied $\sigma$ to 120~{\rm km/s}, and $v_{lag}$ from 0 to 150~{\rm km/s} fixing the density to $\rho_{dd}$ = 0.25$\rho_{H}$.
Figure~\ref{fig:fig4} shows the distributions of combined dark matter halo and dark disc in the solar reference frame. 

Note that for all our studies we have used a local dark matter halo density of $\rho_{H}$ = $0.3~{\rm GeV/cm}^3$. This value is only relevant for the overall normalisation of the capture rate and has no impact on the conclusions drawn from the presented studies. The uncertainty on the local halo density directly relates to the uncertainty in the capture rate and under the assumption of the equilibrium condition to self-annihilation rate. The actual value of local dark matter density is subject to an uncertainty of around a factor of two (see for example~\cite{Green:2011bv} and references therein).

\section{Uncertainties on the capture rate}
\label{sec:uncertainty}

In this section we discuss uncertainties associated with WIMP capture in the Sun. We evaluate the impact of the orbital velocity of the Sun, the halo escape speed, various VDFs for the dark matter halo and the dark disc introduced in the previous section on capture rate, respectively.
For our evaluation we use DarkSUSY~\cite{Gondolo:2004sc}, with which we can precisely compute capture rates and easily modify parameters of the velocity distribution or introduce new VDFs.

\begin{figure*}
\begin{minipage}[t]{.5\textwidth}
\includegraphics[width=1.\textwidth]{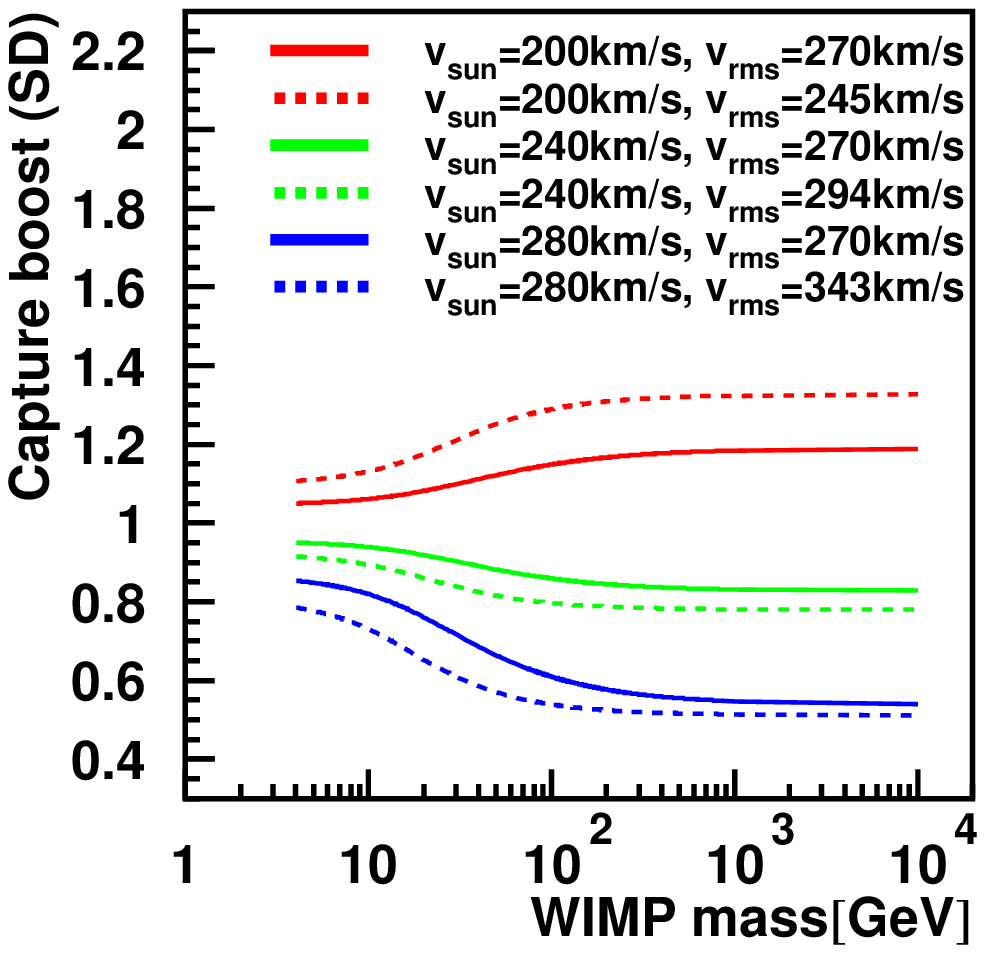}
\end{minipage}
\hfill
\begin{minipage}[t]{.5\textwidth}
\includegraphics[width=1.\textwidth]{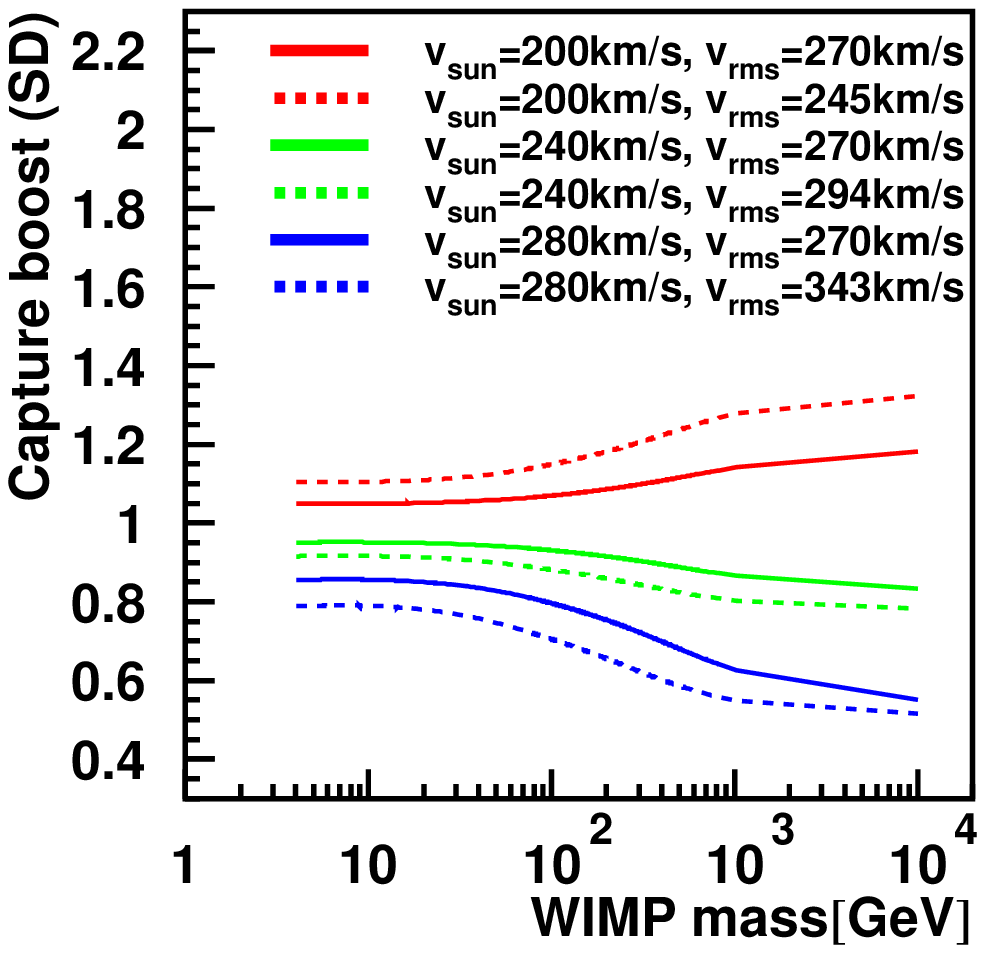}
\end{minipage}
\hfill
\caption{
The capture boosts defined as $C/C_{\rm SHM}$ are shown as function of WIMP mass for the SD~(left) and SI~(right) couplings assuming an orbital speed of the Sun of $v_\odot =$ 200~{\rm km/s} (red solid), 240~{\rm km/s} (green solid) and 280~{\rm km/s}(blue solid) with fixed $v_{rms}=270~{\rm km/s}$, dashed lines for varied $v_{rms}=\sqrt{3/2}v_{\odot}$ with same color scheme.
}
\label{fig:fig5}
\end{figure*}

Figure~\ref{fig:fig5} shows the change in WIMP capture rate in the Sun with respect to the orbital speed of the Sun as function of the WIMP mass. We took the range of 200 $\sim$ 280 km/s for $v_{\odot}$ from McMillan et al.~\cite{McMillan:2009yr} and considered both cases of fixed $v_{rms}$ to 270 km/s and varied $v_{rms}$ to $\sqrt{3/2}v_{\odot}$.
The change in capture rate can best be visualised by showing the relative change compared to the SHM with fiducial value $v_\odot=220~{\rm km/s}$. For this purpose we introduce a boost factor, which is defined by the ratio of the capture rate of the assumed velocity distribution~($C$) divided by the capture rate of SHM~($C_{\rm SHM}$) as $C/C_{\rm SHM}$. Results are separately shown for WIMPs which undergo a SD or SI interaction only as each process experiences different effective capture from the nuclei inside the Sun.
Our results are in agreement with a previous study~\cite{Rott:2011fh} that only considered SD couplings.
Higher $v_{\odot}$ boosts WIMPs in the solar reference frame and makes them harder to be captured. Especially the capture for heavy WIMPs, which is sensitive to the low-velocity region experiences a significant change; the size of the effect appears larger for heavier WIMP, for example for 20/1000~GeV WIMP with SD (SI) coupling the capture rate (solid lines) changes as much as $-24~\%/-45~\%$ ($-15~\%/-38~\%$) when $v_\odot = 280~{\rm km/s}$.
For the case when we varied both $v_{rms}$  and $v_\odot $, the change in capture rate compared to the SHM can be further elevated (dashed lines).

\begin{figure*}
\begin{minipage}[t]{.5\textwidth}
\includegraphics[width=1.\textwidth]{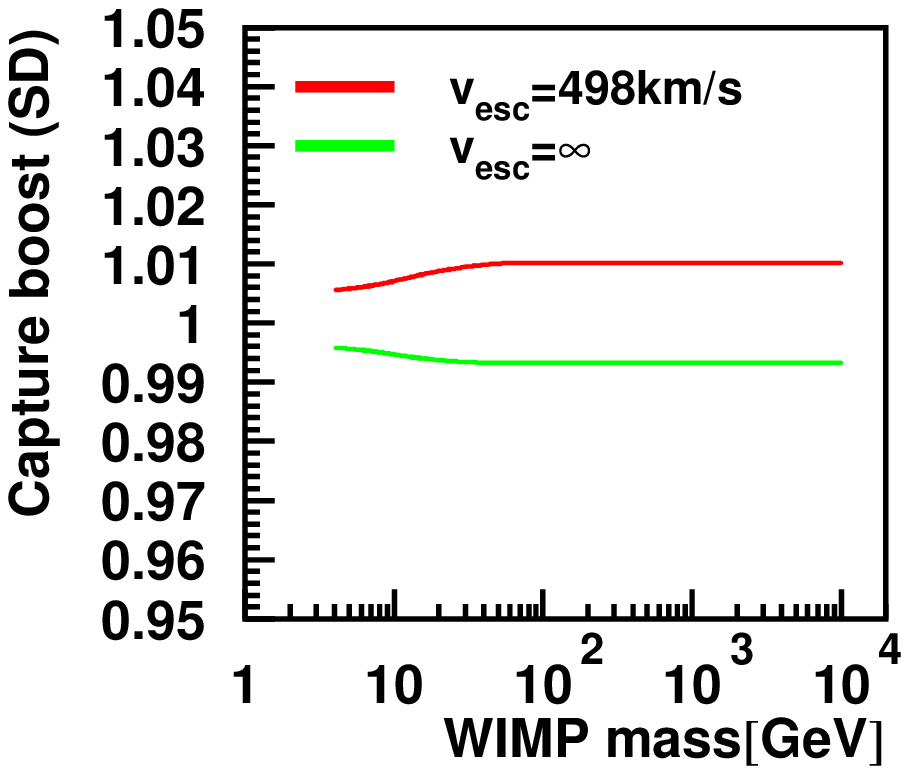}
\end{minipage}
\hfill
\begin{minipage}[t]{.5\textwidth}
\includegraphics[width=1.\textwidth]{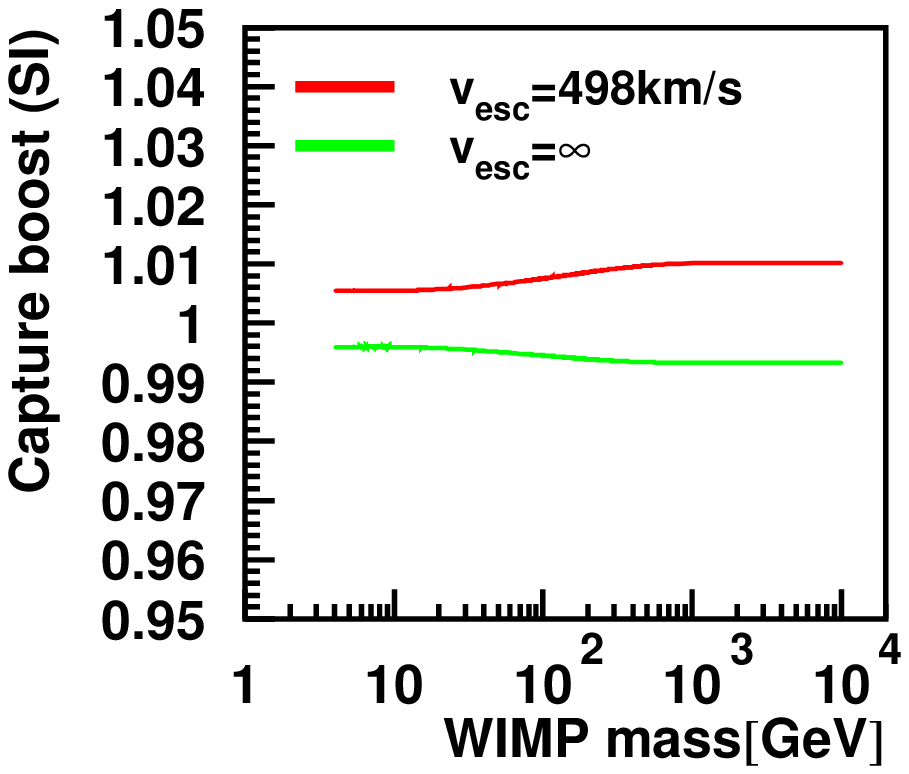}
\end{minipage}
\hfill
\caption{
The capture boosts for SD~(left), SI~(right) coupling WIMPs with the escape speed of the Milky Way of red : 498~{\rm km/s}, green : infinity, compared to 544~{\rm km/s} in SHM as function of WIMP mass. The normalisation of the VDFs are kept to unity for different truncations.}
\label{fig:fig6}
\end{figure*}

High-velocity WIMPs might escape from the Milky Way leaving a cut off in the VDF at the Galactic escape velocity. N-body simulations of Milky Way type halos show a steeper fall-off at the high-velocity tail than expected from a Maxwell-Boltzmann distribution~\cite{Lisanti:2010qx,Strigari:2012up}.
To quantify the change in the capture rate coming from the high-velocity cut we compute the capture boosts for the different cut values compared to our default value of 544~{\rm km/s}. Figure~\ref{fig:fig6} shows the capture boost where the high-velocity cut is decreased to 498~{\rm km/s}~\cite{Smith:2006ym} and increased to be infinity~\cite{Fornasa:2013iaa,McMillan:2011wd}. Indirect detection of heavy WIMPs is blind to the reduction of WIMP population in the high-velocity tail, the marginal changes at high mass come from the renormalization of low-velocity part.
As the fraction of WIMPs in the high-velocity tail is very scarce and kinematically hard to capture, the changes in the low mass region is also rather small. As a result, the change in capture rate is less than 1\% for both SD and SI couplings.

\begin{figure*}
\begin{minipage}[t]{.5\textwidth}
\includegraphics[width=1.\textwidth]{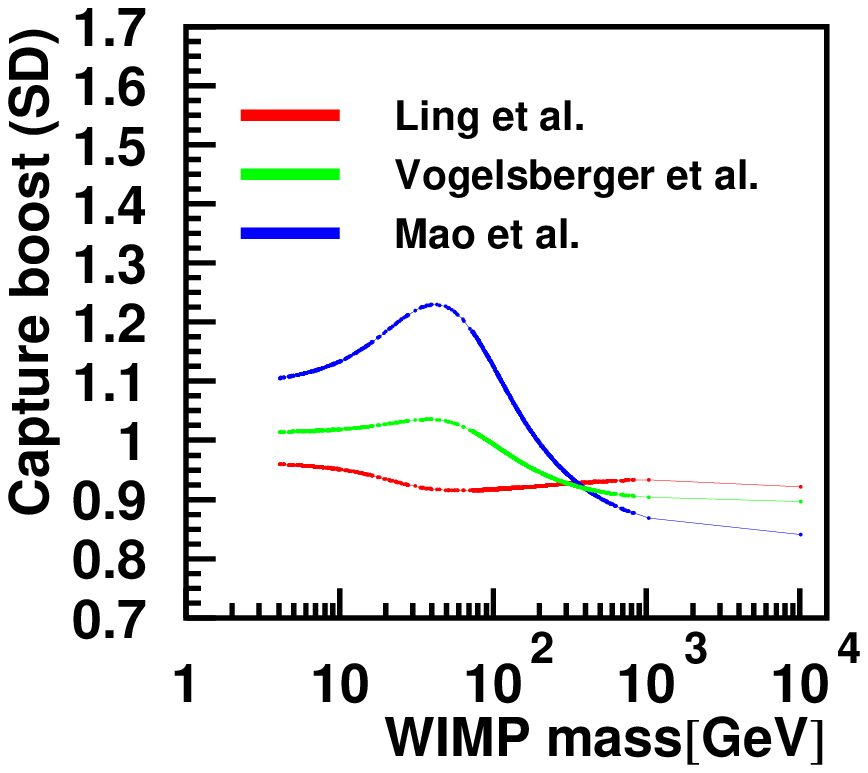}
\end{minipage}
\hfill
\begin{minipage}[t]{.5\textwidth}
\includegraphics[width=1.\textwidth]{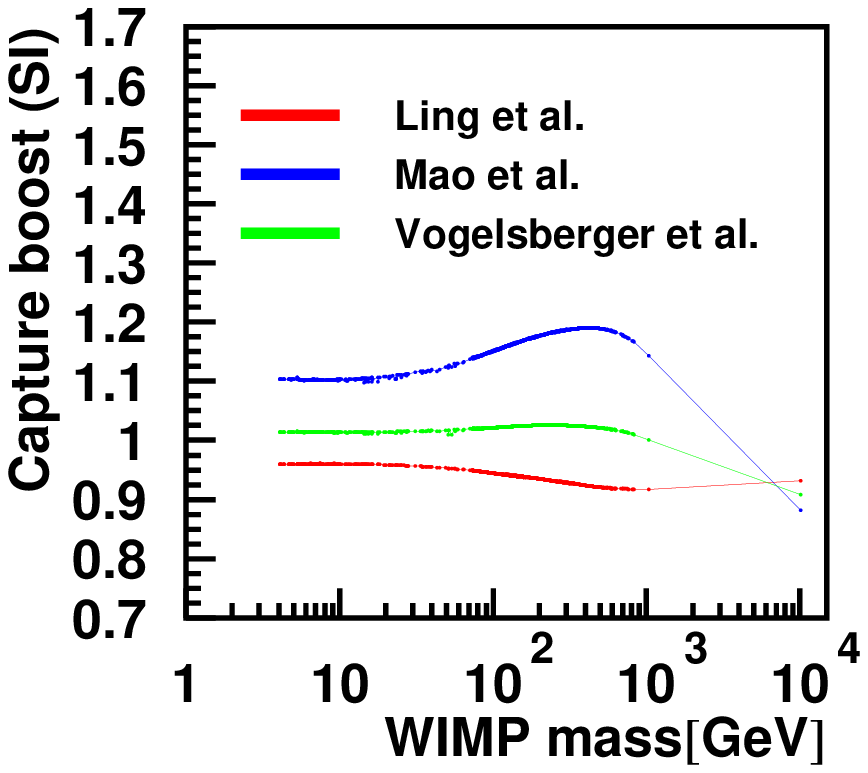}
\end{minipage}
\hfill
\caption{
The capture boosts for SD~(left), SI~(right) coupling WIMPs as function of WIMP mass for VDFs of the dark matter halo in Ling et al.~\cite{Ling:2009eh} (red), Vogelsberger et al.~\cite{Vogelsberger:2008qb} (green) and Mao et al.~\cite{Mao:2012hf} (blue) compared to SHM. Detailed descriptions of VDFs can be found in section~\ref{sec:vdf}.
}
\label{fig:fig7}
\end{figure*}

We now discuss the impact of the VDF itself on the capture rate. In Fig.~\ref{fig:fig7}, we show the relative changes in capture rate for our set of benchmark velocity distributions.
The structure seen in Vogelsberger et al.~\cite{Vogelsberger:2008qb} dark matter halo shown in Fig.~\ref{fig:fig3} (right) is spread out in the local frame by relatively rapid speed of the Sun, seen in Fig.~\ref{fig:fig3} (left). As a result, the SD (SI) coupling capture of 10~GeV WIMPs which comes from entire range of the VDF including the non-smooth shape above 250~{\rm km/s} in Vogelsberger et al.~\cite{Vogelsberger:2008qb} results in a change of $\sim$2 (1)\% (green) compared to the SHM assuming the same local density of WIMPs.
In the same simulation~\cite{Vogelsberger:2008qb}, authors found that the contributions from either unbound streams or sub-halos are likely sub-dominant. Ultra-fine structures beyond the highest resolution of current N-body simulations were found to have negligible impact on direct detection~\cite{Vogelsberger:2010gd,Schneider:2011zk,Fantin:2011nt}. Note that the effect of these properties on the capture rate in the Sun is expected to be more suppressed because of the larger size and the longer time scale of the capture.

Because the capture rate is not sensitive to specific changes in the structure of the VDF, we can simply interpret the overall effect as a shift of the WIMP population from the high- to low-velocity region or vice versa. In Fig.~\ref{fig:fig1}, we show that for 50 (500)~GeV WIMPs the SD (SI) capture process is efficient for velocity below 300~{\rm km/s}. The narrow peak of $f(u)/u$ of the Mao et al.~\cite{Mao:2012hf} halo in this region seen in Fig.~\ref{fig:fig3} (right) brings 23 (19)\% increase (blue) of SD (SI) capture rate for 50 (500) GeV WIMP and keeps positive boost for lighter WIMPs. 
We find that the Ling et al.~\cite{Ling:2009eh} dark matter halo with broad shape and high velocity dispersion gives relatively small abundance in this velocity range, resulting in $\sim$8 (8)\% decrease (red) of the SD (SI) capture rate for a 50 (500)~GeV WIMP and keeps the negative boost for lighter WIMPs.

\begin{figure*}
\begin{minipage}[t]{.5\textwidth}
\includegraphics[width=1.\textwidth]{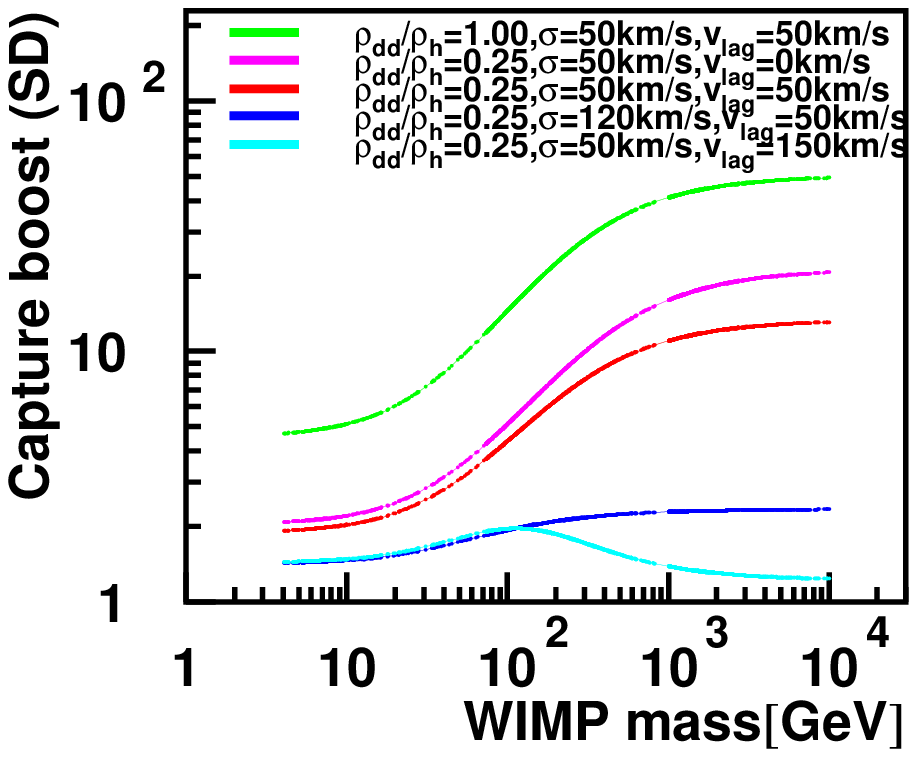}
\end{minipage}
\hfill
\begin{minipage}[t]{.5\textwidth}
\includegraphics[width=1.\textwidth]{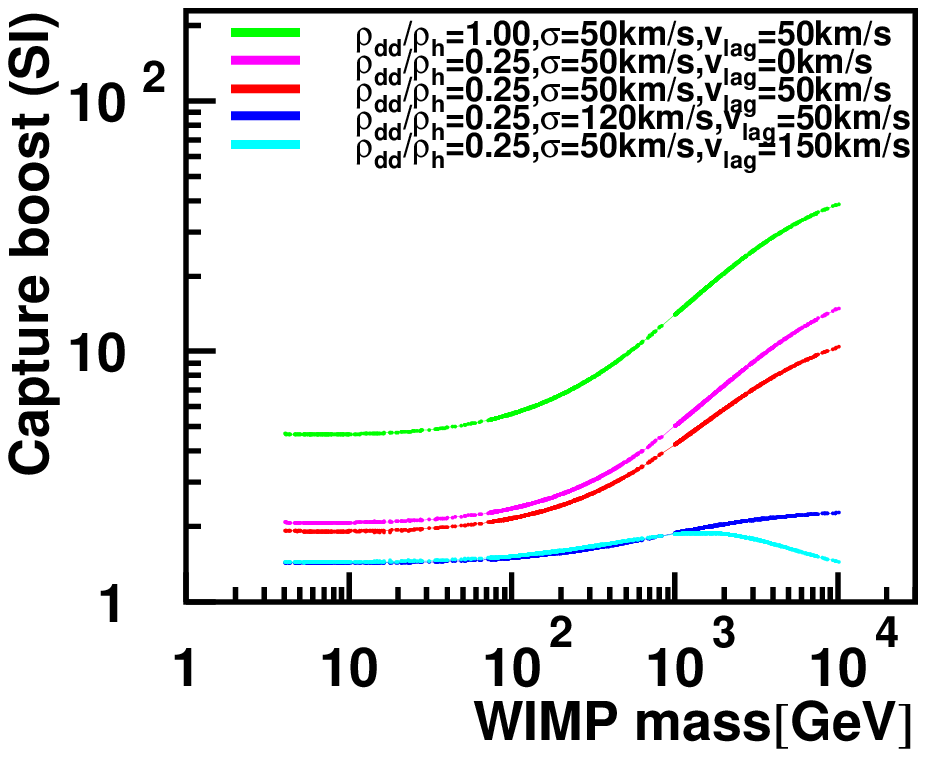}
\end{minipage}
\hfill
\caption{The capture boosts for SD~(left), SI~(right) coupling WIMPs for dark disc scenarios.
green: with $\rho_{dd}/\rho_{H}=1$,
magenta: $v_{lag}=0~{\rm km/s}$,
red: $\rho_{dd}/\rho_{H}=0.25$, $\sigma=50~{\rm km/s}$, $v_{lag}=50~{\rm km/s}$,
blue: $\sigma=120~{\rm km/s}$,
cyan: $v_{lag}=150~{\rm km/s}$.
Unless otherwise specified in the legend, parameters are fixed to $\rho_{dd}/\rho_{H}=0.25$, $\sigma=50~{\rm km/s}$, $v_{lag}=50~{\rm km/s}$ and $\rho_{H}$ = $0.3~{\rm GeV/cm}^3$ (Note that the total dark matter density is increased to $\rho_{dd} + \rho_{H}$). The boosts in capture rate as function of WIMP mass are given relative to a scenario without a dark disc. 
}
\label{fig:fig8}
\end{figure*}

Figure~\ref{fig:fig8} shows the capture boosts for scenarios with a dark disc.
The dark disc is expected to co-rotate with the visible stellar disc and hence their relative velocity with respect to the Sun is smaller compared to that of a non-rotating dark matter halo. Therefore a dark disc will primarily populate the low-velocity phase space from which the Sun can efficiently capture WIMPs.
We consider SI and SD couplings exclusively, which was not done previously. 
First thing we notice is that the boost effect appears more dramatically in SD case. 
The reason for this is that SD capture is more sensitive to the low-velocity region; 
Figure~\ref{fig:fig4} shows that the dark disc contribution mainly is concentrated below 200~km/s. In Fig.~\ref{fig:fig1}, we show that for WIMP masses above 200~GeV (1000~GeV) for SD (SI) coupling, efficient capture occurs for WIMPs with relative velocity to the Sun below 200~{\rm km/s}. Taken together, the boost factor is as large as 15 (6) for SD (SI) coupling at 100~GeV for a massive~($\rho_{dd}/\rho_{H}=1$, green) dark disc model, and can be even larger for more massive WIMPs. 

Our result on SI coupling is in agreement with a previous study done by Ling~\cite{Ling:2009cn}.  A study by Bruch et al.~\cite{bruch:2009rp} has discussed CMSSM models that define the ratio between SI and SD couplings for specific scenarios and found boost factors as much as about an order of magnitude for 100~GeV WIMPs. By running DarkSUSY~\cite{Gondolo:2004sc}, we confirm that CMSSM WIMP models with mass of 100~GeV can result in an order of magnitude boost for a massive~($\rho_{dd}/\rho_{H}=1$) dark disc. We point out that this high boost can be only realised in scenarios with large SD contributions.
Lastly, we confirm that with larger values of relative velocity or velocity dispersion used for modelling the dark disc (blue, cyan), the capture boost can be rather limited~\cite{Ling:2009cn}.

The phase space of low-velocity WIMPs in the solar system could be influenced by the gravitational fields of the planets and have formed a diffused population in the Solar System, becoming a possible source of uncertainty in the low-velocity tail from where the Sun can gravitationally capture WIMPs (known as solar depletion).
After extensive discussions about this effect~\cite{Gould:1990ad,Gould:1999je,Damour:1998rh,Lundberg:2004dn,Peter:2009mi,Peter:2009mk,
Peter:2009mm,Sivertsson:2009nx}, the most recent study~\cite{Sivertsson:2012qj} concluded that solar depletion can be compensated with the loss of the weakly captured population by the Sun due to Jupiter depletion, which has been known as another source of uncertainty for capture of heavy WIMPs.
As a result, the simple free-space approximation in~\cite{Gould:1990ad} is turned out to be accurate~\cite{Sivertsson:2012qj} and planetary effects can be ignored.

\section{Discussions}
\label{sec:discussion}

Direct detection experiments and indirect searches for dark matter captured in the Sun show very different responses to the WIMP velocity distribution.
We first compare the different responses qualitatively and then discuss some specific scenarios and their impact on the detection methods. Our emphasis is on showing the reader how direct and indirect searches complement each other and how the interpretation of results changes for different halo assumptions, and how signals observed in either direct, indirect, or in both methods could be interpreted to understand the Milky Way dark matter halo. For our discussion we distinguish between scenarios with high (above about $100$~GeV) and low WIMP masses.

First, WIMPs in the Sun have been accumulating over a very long time period. Even though there is no evidence for very large density fluctuations or changes in velocity distribution, we point out that direct detection rates are only sensitive to the present local dark matter density and velocity distribution, while indirect detection rates are a result of the dark matter halo density and velocity distribution that has been sampled by the Sun over a time span of the equilibration time scale. Hence, indirect detection rates are related to the average dark matter density and velocity distribution at the solar circle. 

Second, indirect searches for self-annihilating dark matter in the Sun will be sensitive to those parts of the VDF that are efficiently captured. Whereas capture of light WIMPs will occur from wide range of the VDF, heavy WIMPs can only be captured by the Sun from the low velocity part, especially for SD coupling case. In contrast direct detection of heavy WIMPs will generally produce more pronounced signals and as such can sample a wider range of the VDF compared to light WIMPs scenarios which need to be part of the high-velocity tail in order to produce observable recoils. The exact dependence for light WIMP direct detection from the high-velocity tail depends on the material and energy threshold of individual detector. 

When interpreting direct and indirect detection results one needs to keep in mind the contrasting responses of direct and indirect searches to the velocity distribution. 
In light of a source of uncertainty, some of them manifests itself only in the high-velocity tail or the low-velocity distribution. In this case, the complementarity of indirect and direct searches will result in either one of them being insensitive or less impacted compared to the other method, as a result the effects in both detection methods can in many cases be regarded as roughly uncorrelated.
For example in light WIMP scenarios the uncertainty on the high-velocity tail requires a careful interpretation for direct detection experiments~\cite{Ling:2009cn,McCabe:2010zh,Catena:2011kv,Lisanti:2010qx,Bhattacharjee:2012xm}, while the size of the effect on capture rate in the Sun as studied here will be marginal. Note that several studies have pointed out the large uncertainty in modeling of the high-velocity tail of the dark matter halo~\cite{Lisanti:2010qx,Catena:2011kv,Kundu:2011ek,Green:2011bv}, which merits the complementarily of direct and indirect searches. In a heavy WIMP scenario the opposite effect will occur; indirect detection results need to be interpreted in careful manner as the capture will be governed by the low-velocity part of the VDF. Luckily, sources of uncertainties in the low-velocity tail such as the shape of simulated halos or bound WIMP population in the Solar System are expected to be small~\cite{Sivertsson:2012qj,Arina:2013jya}.

As another example of uncertainties which will change the VDF in uncorrelated manner is the existence of a dark disc. The additional population of WIMPs in the low-velocity region from the dark disc will significantly enhance heavy WIMP capture and therefore indirect detection rates, whereas for direct detection experiments considerable~\cite{Frandsen:2011gi,Billard:2012qu,Fairbairn:2012zs}, but smaller effects are expected (up to factor three for WIMP masses above 50 GeV~\cite{Bruch:2008rx} for recoil energies of 5 $-$ 20 keV). For lower masses indirect detection rates could experience sizable boosts depending on the dark disc scenario, however direct detection rates will be hardly altered. The annual modulation signal in the so-called channeling region of DAMA/Libra was studied and found not to be changed~\cite{Ling:2009cn}. 
The addition of a dark disc will not change or reduce the original dark matter halo but only gives an extra population to which indirect searches are significantly more sensitive than direct searches. Hence, constraints from indirect searches assuming SHM are more conservative with respect to direct searches if one considers the existence of a dark disc.

Some changes in the halo models or astrophysical assumptions will give rise to an overall shift of the WIMP population from the high-velocity to the low-velocity region or vice versa. In this case, the expected change in event rates for both direct and indirect searches are rather moderate. For the orbital speed of the Sun, the uncertainty is found to be less than $\sim$40\% in this work and $\sim$20\% for various direct detector set-ups examined~\cite{Serpico:2010ae}. The expected change in capture rate due to the halo VDF is found to be less than $\sim$20\% in this work and a similar magnitude is expected for direct detection experiments in general~\cite{Vogelsberger:2008qb,Green:2010gw,Serpico:2010ae,Catena:2011kv}. The effect is smaller in direct detection in heavy WIMP scenarios and for indirect detection for light WIMP scenarios, as for those cases a wide range of the velocity distribution contributes. For example, even if WIMP distribution has shifted from the low-velocity to the high-velocity region which is harder to be captured, they can still contribute to the indirect neutrino signal in a light WIMP scenario, however not in a heavy WIMP scenario.

The responses from indirect and direct searches to uncertainties resulting in an overall shift of the velocity distribution can often manifest themselves in an anti-correlated way; i.e. where the capture rate is boosted due to an increased low-velocity population, direct detection rates could be decreased due to a deficit in the high-velocity population, and vice versa.
As an example, a higher orbital speed of the Sun enriches the high-velocity region and could increase signal rates for direct searches~\cite{McCabe:2010zh}, resulting in an opposite change~\cite{Serpico:2010ae} compared to indirect searches, as shown in this work. 
A similar effect occurs for our adopted halo benchmark model of Mao et al.~\cite{Mao:2012hf} which leads to a reduction in direct detection rates for 10~GeV WIMPs due to the depletion of the VDF region above 300~{\rm km/s} ($u_{min}/u_{esc} > 0.55$, see the blue line in Fig.6 of Mao et al.~\cite{Mao:2012hf}), however in contrast the capture in the Sun would be increased as found in this work.
While the different response to halo uncertainties is a limiting factor in the comparison between direct and indirect detection~\cite{Serpico:2010ae} it also stresses the complementarity between the two methods. 

Overall, the fact that even extreme cases of the VDF do not change indirect detection sensitivities dramatically and the only exception to this, a dark disc, would boost rates significantly, leaves indirected searches to be conservative in comparison to direct searches.

\section{Conclusions}
\label{sec:Conclusions}

The dark matter phase distribution is a large source of uncertainty for WIMP searches. Quantifying the size of uncertainties on the WIMP capture rate based on the velocity distribution is of extreme importance to interpret results from indirect searches for WIMP annihilation in the Sun together with direct searches. In our detailed study we have evaluated the capture rate of WIMPs in the Sun in dependence on the velocity distribution. We have quantified uncertainties as function of the WIMP mass and considered various solar circular velocities, a high-velocity cut-off, VDFs motivated by various simulations as well as scenarios with a dark disc.

We conclude that uncertainties coming from parameters of the dark matter velocity distribution are not very significant; For a WIMP mass of 20~GeV, we find an uncertainty of 18 (14)\% from local circular speed, 1 (1)\% from high-velocity cut for the tested range for SD (SI) couplings. We have further shown that while the dark matter halo VDF itself is unknown the impact on the capture rate of light WIMPs is suppressed because the effect is integrated over a large velocity range, which is consistent with the discussion in~\cite{Wikstrom:2009kw}; it is up to 16 (10)\% for the dark matter halos from cosmological simulations.
In summary, the estimated size of the uncertainty from possible sources in VDF is found to be less then 24\% (17\%) for 20~GeV SD (SI) coupling WIMPs, for scenarios without a dark disc.

For heavier WIMPs, the uncertainties are larger but still below about 50\%; they are 45 (32)\% from local circular speed, 1 (1)\% from high-velocity cut for the tested range, up to 11 (16)\% from deviation of the VDF from Maxwellian for SD (SI) coupling 500 GeV WIMPs.

Through detailed studies of the individual uncertainties on the dark matter capture rate, we conclude that the overall uncertainties in indirect searches are moderate. Interpretation of indirect results typically carries an error smaller than 50\%. An existence of dark disc enriching low-velocity population could result in a significant increase in capture rate especially for heavy WIMPs, but it will enhance the capture rate therefore allow the interpretation of current limits from indirect searches to be more conservative.

We further emphasis that uncorrelated and anti-correlated sensitivities of direct and indirect searches to the high- and the low-velocity tails can potentially be used to disentangle particle physics from astrophysical assumptions and their uncertainties. Indirect detection of light WIMPs can be a good complementary method to break degeneracies badly afflicting the interpretation of direct detection experiments~\cite{Pato:2010zk}. For a high mass WIMP scenario, sensitivity of indirect searches to the low-velocity region can be exploited to understand properties such as a dark disc component to which direct detection experiments are rather insensitive~\cite{Peter:2013aha}.
In terms of not only good sensitivity for signal but also unique sensitivities and insensitivities to astrophysical properties, proposed large volume low threshold neutrino detectors such as the Hyper-Kamiokande~\cite{Kearns:2013lea} and the PINGU upgrade~\cite{Aartsen:2014oha} to IceCube~\cite{Aartsen:2012kia} are expected to play important roles in future searches.
Finally, we note that for reliable usage of results from indirect detection experiments, the equilibrium condition between capture and annihilation in the Sun needs to be evaluated for specific WIMP scenarios in individual studies~\cite{Ellis:2009ka,Rott:2011fh,Serpico:2010ae}.

\begin{acknowledgments}
This work was supported through the CCAPP visitor program and by the GCOE Program QFPU of Nagoya University from JSPS and MEXT of Japan. CR acknowledges support from the Basic Science Research Program through the National Research Foundation of Korea funded by the Ministry of Education, Science and Technology (2013R1A1A1007068). We thank Matthias Danninger, Annika Peter, and Mark Vogelsberger for discussions and comments on the manuscript. We would like to thank the referee for useful suggestions and comments.
\end{acknowledgments}

\bibliographystyle{JHEP}
\bibliography{JCAP_Choi_Rott_Itow}

\end{document}